\documentclass[prd,tightenlines,floatfix,showpacs,preprintnumbers,nofootinbib,eqsecnum]{revtex4}
 \usepackage[dvips,final]{graphicx}
  \usepackage{amssymb}
   \usepackage{amsmath} % Advanced math typesetting
    \usepackage{amsfonts}
     \usepackage{epsfig}
     \usepackage{color}
      \usepackage{bm} % bold math
        \usepackage{xcolor}
        
\usepackage{booktabs}
\usepackage{multirow}
\usepackage{slashed}
\usepackage{comment}

\def\doublespace{\def\baselinestretch{1.6}\large\normalsize}
\def\normalspace{\def\baselinestretch{1.0}\normalsize}
\def\PSfig#1#2{\scalebox{#1}{\includegraphics{#2}}}

\def\Caption#1{
  \normalspace
  \vskip-1mm\caption{\sl#1}\vskip-1mm
  \doublespace
}

\newcommand\la{\langle}
 \newcommand\ra{\rangle}
 \newcommand\beq{\begin{equation}}
 
 \newcommand\eeq{\end{equation}}
 \newcommand\beqn{\begin{eqnarray}}
 \newcommand\eeqn{\end{eqnarray}}

\def\fm{\,\mbox{fm}}
\def\MeV{\,\mbox{MeV}}
\def\GeV{\,\mbox{GeV}}
\def\TeV{\,\mbox{TeV}}

\begin{document}
%%%%%%%%%%%%%%%%%

%================================================
\title{
\vspace*{-2.0cm}
Unconventional mechanisms 
of heavy quark fragmentation
%
%Heavy quarkonium production \\
%in ultra-peripheral nuclear collisions
}
%================================================

\author{B. Z. Kopeliovich$^1$}
\email{boris.kopeliovich@usm.cl}

\author{J. Nemchik$^{2,3}$}
\email{jan.nemcik@fjfi.cvut.cz}
%\email{nemcik@saske.sk}

\author{I. K. Potashnikova$^1$}
\email{irina.potashnikova@usm.cl}

\author{Ivan Schmidt$^1$}
\email{ivan.schmidt@usm.cl}

\vspace*{0.5cm}

\affiliation{
\vspace*{0.2cm}
$^1$Departamento de F\'{\i}sica,
Universidad T\'ecnica Federico Santa Mar\'{\i}a,\\
Avenida Espa\~na 1680, Valpara\'iso, Chile}
\affiliation{$^2$
Czech Technical University in Prague, FNSPE, \\B\v rehov\'a 7, 11519
Prague, Czech Republic}
\affiliation{$^3$
Institute of Experimental Physics SAS,\\ Watsonova 47, 04001 Ko\v sice, Slovakia
}

\vspace*{2.0cm}
\date{\today}
%%%%%%%%%%%%%%%%%%%%%%%%%
\begin{abstract}
%%%%%%%%%%%%%%%%%%%%%%%%%
\vspace*{5mm}
Heavy and light quarks produced in high-$p_T$ partonic collisions radiate differently.
Heavy quarks regenerate their color field,  stripped-off in the hard reaction, much faster than the light ones
and radiate a significantly smaller fraction of the initial quark energy.
This peculiar feature of heavy-quark jets leads to a specific shape of the fragmentation functions observed in $e^+e^-$ annihilation. Differently from light flavors, the heavy quark fragmentation function strongly   peaks at large fractional momentum $z$, i.e. the produced heavy-light mesons, $B$ or $D$, carry the main fraction of the jet momentum. This is a clear evidence of the dead-cone effect, 
and of a short production time of a heavy-light mesons.
Contrary to propagation of a small $q\bar q$ dipole, which survives in the medium due to color transparency, a heavy-light $Q\bar q$ dipole promptly expands to a large size. Such a big dipole has no chance to remain intact in a dense medium produced in relativistic heavy ion collisions. On the other hand, a breakup of such a dipole does not affect much the production rate of $Q\bar q$ mesons, differently from the case of light $q\bar q$ meson production.%%%%%%%%%%%%%%%%%%%%%%%%
\end{abstract}
%%%%%%%%%%%%%%%%%%%%%%%%


\maketitle
\section{Introduction}

\begin{figure}[hbt]
\begin{center}
\includegraphics[width=4.5 cm]{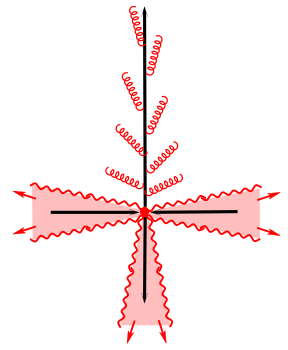}
\caption{\label{fig1} High-$p_T$ collision in the c.m. frame of two partons, which leads to production of four jets: (i)-(ii) soft color field shaken off in the collision; (iii)-(iv) transverse cones of gluons radiated due to regeneration of the stripped off color field.
}
\end{center}
\end{figure}

High-$p_T$ parton scattering leads to 
formation of four cones of gluon radiation:
(i)-(ii) backward-forward jets formed by the color field of the colliding partons 
shaken off in in the hard collision;
(iii)-(iv) the scattered partons carry
 no field up to transverse momenta $k_T<p_T$.
These partons 
are regenerating the lost color field via gluon radiation 
forming the up-down jets, as is illustrated in Fig.~\ref{fig1}
%%%%%%%%%%%%%%%%%%%%%%%
The radiation process is ordered in time or path length according to \cite{lp},
\beq
l_c=\frac{2Ex(1-x)}{k_T^2+x^2m_q^2}\ .
\label{lc}
\eeq
Here $x$ is the fractional light-cone momentum of the radiated gluon; $k_T$ is its transverse momentum relative to the initial quark direction. The radiated gluons subsequently hadronize forming a jet of hadrons. For heavy quarks the second term in the denominator play important role leading to the so called dead-cone effect \cite{troyan}.

In terms of the Fock state representation all radiated
gluons pre-exist in the initial
bare parton, and are liberated  on mass shell successively in 
accordance with their coherence length/time Eq.~(\ref{lc}).
First  are radiated gluons with small longitudinal and large transverse momenta.

\subsection{Radiational energy loss in vacuum}

How much energy is radiated over the path length L? 
Only gluons with radiation length $l_c<L$ contribute \cite{similar},
\beq
\Delta E_{rad}(L) =
\int\limits_{\lambda^2}^{Q^2}
dk_T^2\int\limits_0^1 dx\,\omega\,
\frac{dn_g}{dx\,dk_T^2}\,
\Theta(L-l_c),
\label{130}
\eeq
 where $\omega$ is the gluon energy; the soft cut-off parameter $\lambda=0.2\GeV$.
The perturbative radiation spectrum reads,
\beq
\frac{dn_g}{dx\,dk_T^2} =
\frac{2\alpha_s(k_T^2)}{3\pi\,x}\,
\frac{k_T^2[1+(1-x)^2]}{[k_T^2+x^2m_Q^2]^2}\,.
\label{145}
\eeq
We see that radiation by light and heavy quarks behave quite differently at small $k_T$:\\
(i) Light quarks: $dn_g/dk_T^2\propto 1/k_T^2$\\
(ii) Heavy quarks: $dn_g/dk_T^2\propto k_T^2/m_Q^2$

\begin{figure}[hbt]
\begin{center}
\includegraphics[width=8 cm]{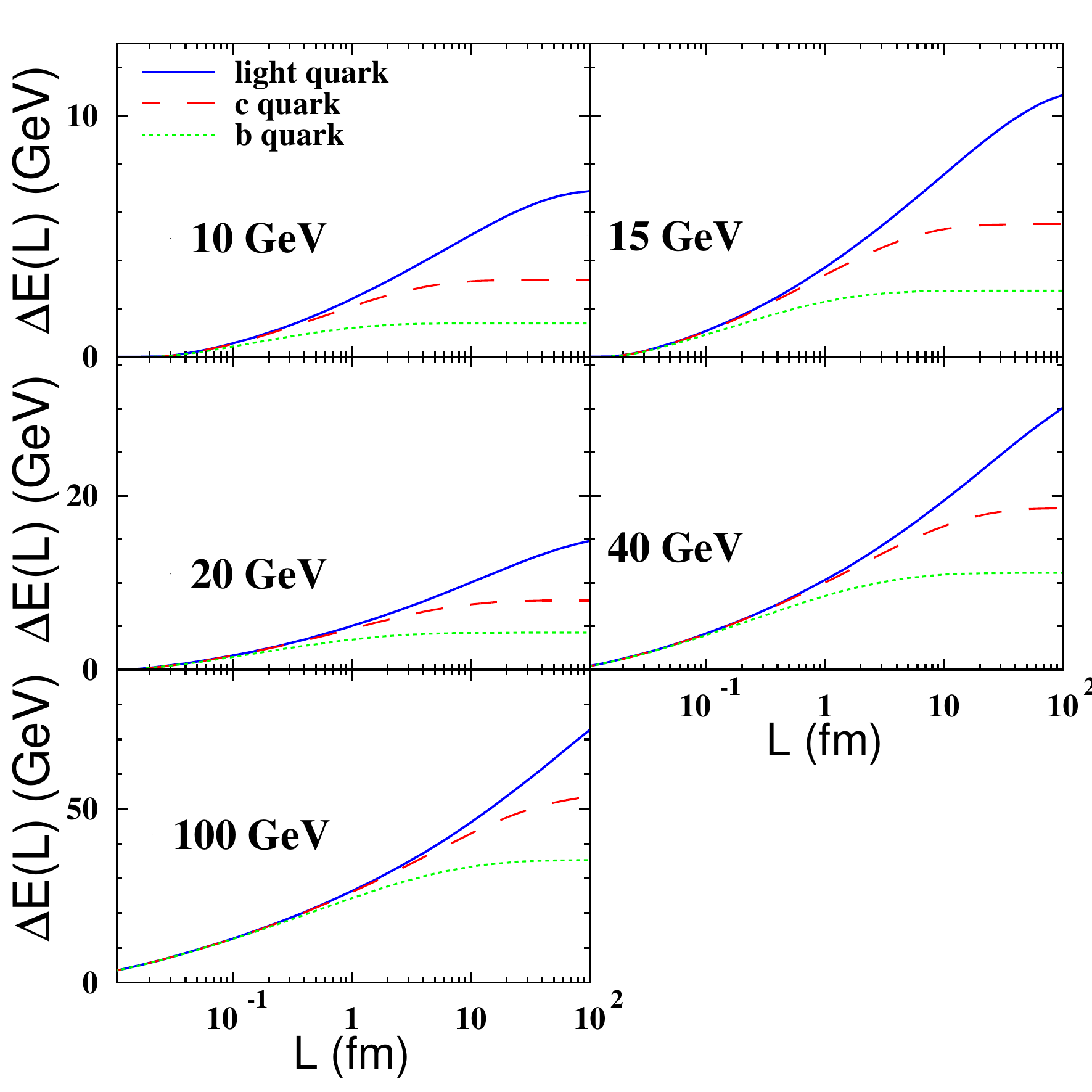}
\caption{\label{dE-E} Radiational energy loss in vacuum by light (u,d), c and b quarks, depicted by blue, red and green curves respectively. Radiated energy $\Delta E$ is plotted as function of path length for different jet energies.
}
\end{center}
\end{figure}

\begin{figure}[hbt]
\begin{center}
\includegraphics[width=8.5 cm]{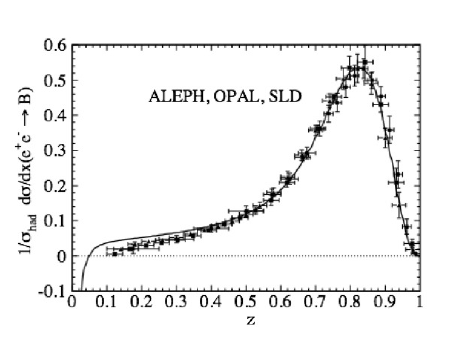}
\caption{\label{ff} The $b\to B$ fragmentation function, 
         from $e^+e^-$ annihilation at LEP. 
         The curve is the DGLAP fit \cite{bottom}.
}
\end{center}
\end{figure}

Dead-cone effect: gluons with $k_T^2<x^2m_Q^2$ 
are suppressed \cite{troyan,similar}. Heavy quarks radiate less energy
compared with the light ones. They promptly restore their color
field and  stop radiating. The amount of radiated energy for light and heavy flavors is depicted in Fig.~\ref{dE-E} vs radiation length for different jet energies.
%%%%%%%%%%%%%
We see that heavy quarks radiate only a small fraction
10-20\% of their initial momentum. In particular, this explains the unusual shape of the
experimentally observed fragmentation function $D_{b/B}(z)$ of b-quarks, presented in Fig.~\ref{ff} \cite{bottom} (and similar for charm \cite{charm}).
%%%%%%%%%%%%%
%%%%%%%%%%%%
Indeed, most of $B$-mesons carry a large fraction 
$z\sim 80\%$, of the $b$-quark momentum.

We conclude that such a specific shape of the fragmentation
function of heavy quarks is a direct manifestation of the dead-cone effect.

\section{Production length}

\begin{figure}[hbt]
\begin{center}
\includegraphics[width=7 cm]{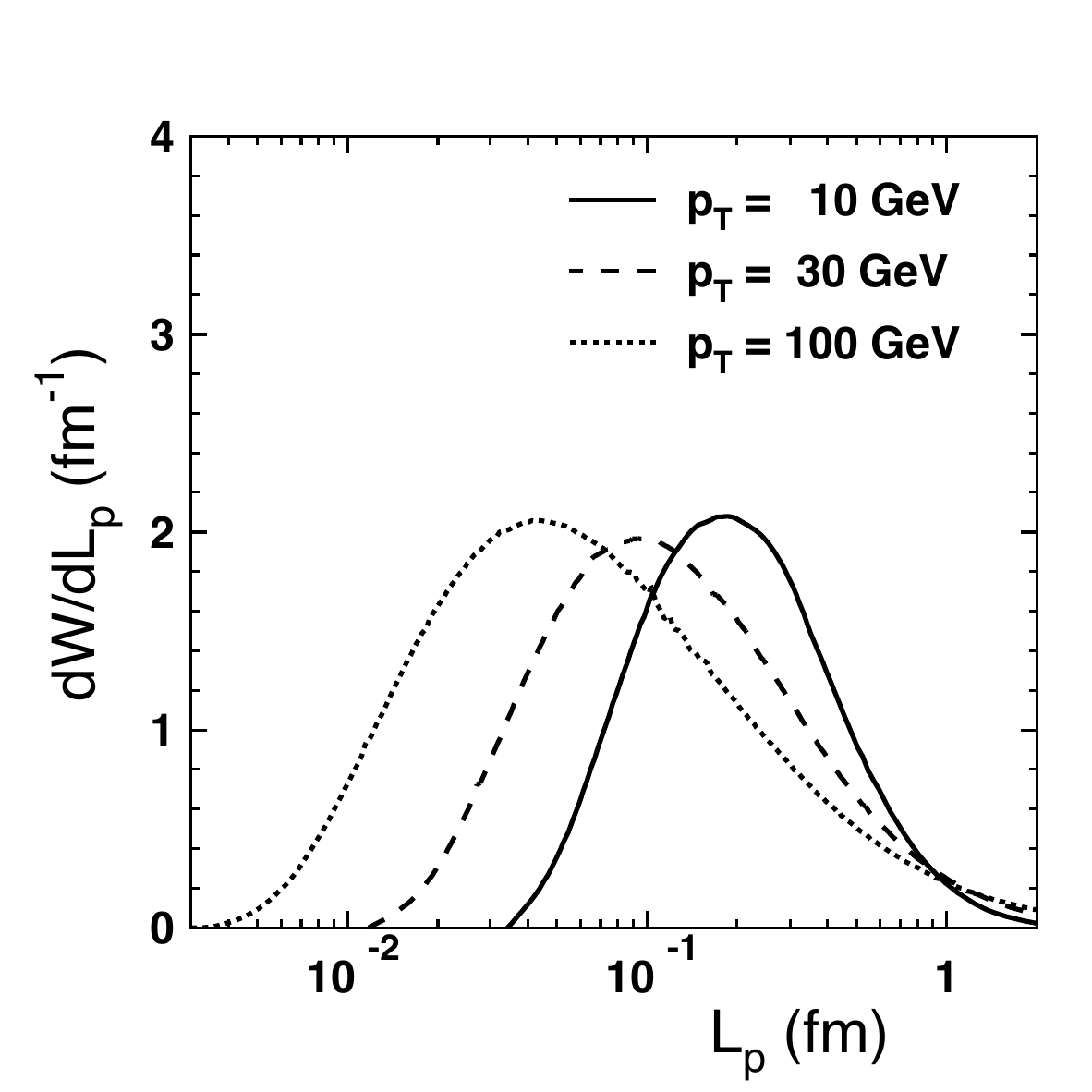}
\caption{\label{Log} The $L_p$-distribution of $B$-mesons produced 
         with different $p_T$ in $pp$ collisions.
}
\end{center}
\end{figure}

The process of gluon radiation by a heavy quark $Q$ ends up
with color neutralization by a light antiquark and
production of a $Q\bar q$ dipole.
As far as we are able to calculate the radiated fraction of the light-cone momentum (e.g. for $b$-quark) $\Delta p_+^b(L)/p_+^b$, the production length $L_p$ distribution $W(L_p)$ can be extracted directly from data on $D_{b/B}(z)$,
\beq
\frac{dW}{dL_p}=
\left.\frac{\partial \Delta p_+^b(L)/p_+^b}{\partial L}\right|_{L=L_p}
\!D_{b/B}(z)\, ,
\label{155}
\eeq
The results for the differential distribution $dW/dL_p$ are depicted in Fig.~\ref{Log} at several values of momenta $p_T$.
%%%%%%%%%%%%%
%%%%%%%%%%%%

Remarkably, the mean value of $L_p$ is extremely short
and shrinks with rising $p_T$. This sounds counter-intuitive,
however, the process has maximal hard scale allowed by the kinematics $p_T=E_{c.m.}/2$.

The production length $L_p$ turns out to
be much shorter than the confinement radius,
indicating that the fragmentation mechanism is pure perturbative.
At $L=L_p$, a small-size dipole $b\bar q$ is produced, with no certain mass, but with a certain radius. It is to be projected on the $B$-meson wave function, giving $\Psi_B(0)$ (compare with \cite{berger}).

\section{Fragmentation in a dense medium}

\subsection{Formation length of a $Q\bar q$ meson}

The light antiquark in the B-meson carries a tiny fraction of its momentum, $x\approx m_q/m_Q$, i.e. about 5\%.
The produced $b\bar q$ dipole has a small transverse separation, but it expands with a high speed, enhanced by 1/x, i.e. is an order of magnitude faster than symmetric $\bar qq$ or $\bar QQ$ dipoles. 
\beq
l_f\sim {1\over2}x(1-x)\la r_T^2\ra p_T,
\label{lf}
\eeq
where $\la r_T^2\ra = {8/3}\la r_{ch}^2\ra$, and $\la r_{ch}^2\ra_B=0.378\fm^2$ as was evaluated in the potential model \cite{radius}. The $B$ meson is nearly as big as the pion, since its radius is controlled by the mass of the light antiquark.

\begin{figure}[hbt]
\begin{center}
\includegraphics[width=13 cm]{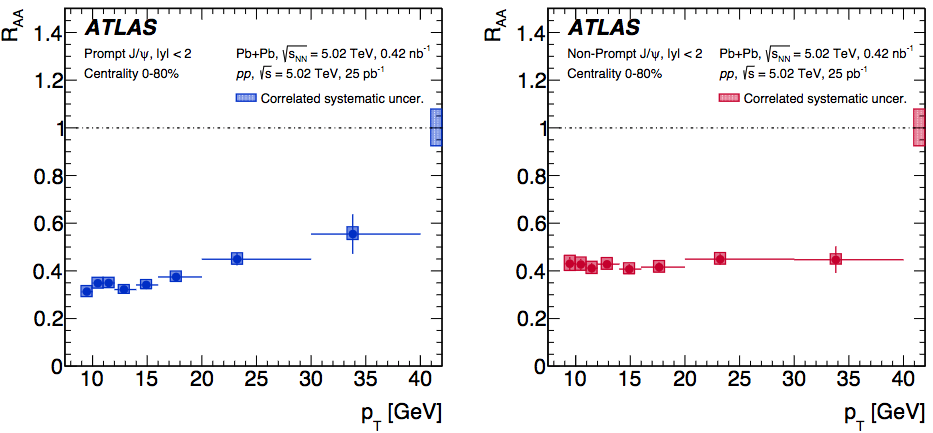}
\caption{\label{ATLAS} Nuclear suppression factor $R_{AA}$ vs $p_T$.
{\it Left}: promptly produced $J/\psi$'s exhibit color transparency effect.
{\it Right}: $J/\psi$'s from $B$ decays demonstrate a $p_T$-independent color-opacity 
effect.}
\end{center}
\end{figure}

According to (\ref{lf}) the dipole heavy-light $Q\bar q$ dipole separation promptly reaches the large hadronic size. This is confirmed by comparison data, for $J/\psi$ detected in $Pb-Pb$ nuclear collisions.
Data demonstrate a color opacity for $B$-mesons (prompt production) 
and color transparency effect for $J/\psi$ decaying to $B$ (non-prompt production). The nuclear suppression factors $R_{AA}$ for these two channels are compared in Fig.~\ref{ATLAS} \cite{atlas-b}.
%%%%%%%%%%%%%
%%%%%%%%%%%%

While Eq.~(\ref{lf}) describes the early, perturbative stage of the dipole expansion,
the further evolution filters out the states with large relative phase shifts.
The longest time takes discrimination between the two lightest hadrons,
the ground state $B$ and the first radial excitation $B^\prime$, which concludes the
formation process. Correspondingly, the full formation path length can be evaluated as,
\beq
l_f=\frac{2p_T}{m_{B^\prime}^2-m_B^2}.
\label{Lf}
\eeq
E.g. for the oscillatory potential $m_{B^\prime}-m_B=0.6\GeV$, so $l_f=0.06\fm[p_T/1\GeV]$ is extremely short for medium-large transverse momenta.
 
 \subsection{Attenuation of dipoles propagating in a dense medium}
 
 The mean free path of a $Q-\bar q$ meson in a hot medium characterizing by the transport coefficient (the rate of broadening) $\hat q$,
 \beq
 \lambda_{Q\bar q}\sim\frac{1}{\hat q\la r_T^2\ra} = \frac{3}{8\hat q\la r_{ch}^2\ra_{Q\bar q}}.
 \label{lambda}
 \eeq
 E.g. at $\hat q=1\GeV^2/\fm$ $\lambda_B=0.04\fm$, so a formed $B$-meson breaks up in the medium nearly instantaneously.
 
 A $b$-quark propagating through the hot medium, easily picks up and loses accompanying light antiquarks without an essential reduction of its momentum. Meanwhile the $b$-quark keeps dissipating its energy with a rate, slightly enhanced by medium induced radiative energy loss \cite{dk} 
effects. Eventually the detected $B$-meson is produced in the dilute periphery of the medium.

\begin{figure}[hbt]
\begin{center}
\includegraphics[width=4 cm]{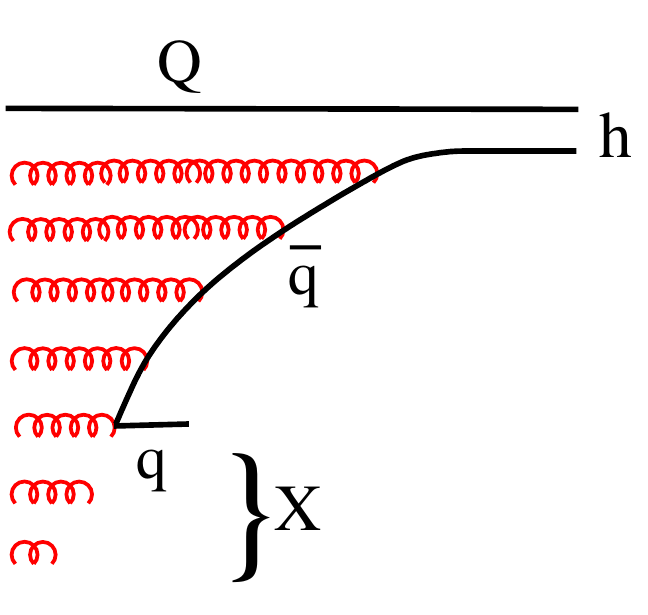}
\caption{\label{regge-cut} Redistribution of energy inside the $Q\bar q$ dipole. The gluons radiated by $Q$ are absorbed by $\bar q$ so the dipole energy remains unchanged.
}
\end{center}
\end{figure}

The heavy quark keeps losing energy even inside a colorless $Q\bar q$   dipole sharing its momentum with the light quark, as is illustrated in Fig.~\ref{regge-cut} presenting a unitarity cut of a $\bar qq$ Reggeon,
 %%%%%%%%%%%%%
%%%%%%%%%%%%
Thus, the heavy quark $Q$ dissipates a part of its energy on a long path from the hard collision point to the medium periphery.
\beq
\frac{dE}{dL}=\frac{dE_{rad}}{dL}-\kappa(T),
\label{e-loss}
\eeq
where $\kappa(T)$ is temperature dependent string tension in the medium \cite{string}
$\kappa(T)=\kappa_0(1-T/T_c)^{1/3}$; the vacuum string tension $\kappa_0=1\GeV/\fm$; The critical temperature is fixed at $T_c=200\MeV$.

\subsection{Medium modified production rate}

The cross section of a heavy-light meson $M$ production in $pp$ collisions can be presented in the factorized form,
\beq
\frac{d^2\sigma_{pp\to M}}{d^2p_T}=\frac{1}{2\pi p_T E_T}
\int d^2 q_T\,\frac{d^2\sigma_{pp\to Q}}{d^2q_T}
\int\limits_0^\infty dL_p \frac{dW}{dL_p}\,\frac{\Delta E(L_p)}{E}\,
\delta\left(1-z-\frac{\Delta E(L_p)}{E}\right)
%z\,D_{Q/M}(z),
\label{890}
\eeq
We replaced the $b\to B$ fragmentation function by the differential expression (\ref{155}).
The medium-modified $L_p$ distribution is given by,
\beq
\frac{dW^{AA}}{dL_p}={1\over2}\la r_B^2\ra \hat q(L_p)\,
\exp\left[-{1\over2}\la r_B^2\ra\int\limits_{L_p}^\infty dL\,\hat q(L)\right]
\label{WAA}
\eeq
Here, for the sake of simplicity, we fixed the $Q\bar q$ dipole separation at the mean value. This approximation is rather accurate due to shortness of $l_f$. Otherwise, one can calculate the attenuation factor in (\ref{WAA}) exactly, applying the path integral technique \cite{kst1,kst2}.

Eventually, the production rate of heavy-light mesons in $AA$ collisions with impact parameter $\vec s$ reads,
\beq
\int d^2 q_T\,\frac{d^2\sigma_{pp\to Q}}{d^2q_T}
\int d^2\tau\, T_A(s)T_A(\vec s-\vec\tau)
\int\limits_0^\infty dL_p \frac{dW^{AA}}{dL_p}\,\frac{\Delta E(L_p)}{E}\,
\delta\left(1-z-\frac{\Delta E(L_p)}{E}\right)
\label{sigmaAA}
\eeq

The effective production length $\tilde L_p$ in the medium turns out to be much longer than in vacuum, because the heavy-light meson is produced mainly at the medium periphery, long distance from the hard collision point.

\subsection{Data analysis}

Now we are in a position to calculate the nuclear ratio
\beq
R_{AA}(\vec{s},\vec p_T) 
=
\frac{d^2\sigma_{AA}(s)/d^2p_Td^2s}
{T_{AA}(s)\,d^2\sigma_{pp}/d^2p_T},
\label{850}
\eeq
to be compared with data. Here
\beq 
T_{AA}(s)=\int d^2\tau T_A(\tau) T_A(\vec s-\vec\tau),
\label{TAA}
\eeq
and $T_A(s)$ is the nuclear thickness function.

The model cannot fully predict (as well as any other model) the nuclear ratio, because the medium density is not known, but is rather the goal of the research. We embedded this information into the broadening rate (transport coefficient) following the popular model \cite{wang}
 \beq
\hat q(l,\vec s,\vec\tau,\phi)=\frac{\hat
q_0\,t_0}{t}\, \frac{n_{part}(\vec s,\vec\tau + l\,\vec p_T/p_T)}{n_{part}(0,0)}
\,\Theta(t-t_0)\, ,
\label{900}
\eeq
where 
 $n_{part}(\vec s,\vec\tau)$ is the number of participants at transverse coordinates
 $\vec s$ and $\vec \tau$ relative to the centers of the colliding nuclei.
 The falling time dependence, $1/t$ is due to longitudinal expansion of the produced medium. 
 The time interval $t_0$ required for equilibrated medium production. We fixed it at the frequently used value $t_0=1\fm$.

\begin{figure}[hbt]
\begin{center}
\includegraphics[width=13 cm]{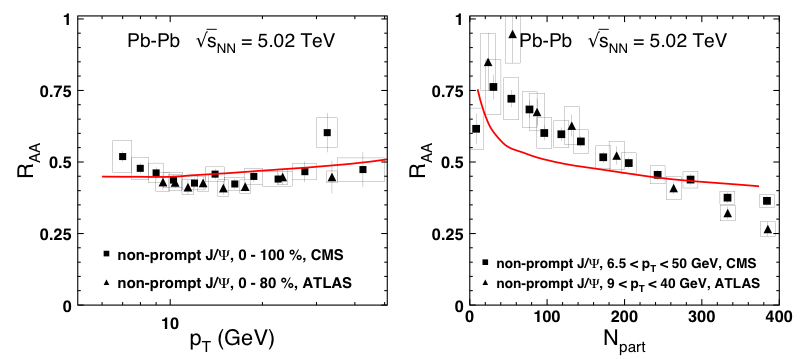}
\caption{\label{Bmeson} Nuclear ratio $AA/pp$ for $B$-meson production in lead-lead collisions as function of $p_T$ ({\it Left}) and 
versus centrality ({\it Right}).
Data are from ATLAS \cite{atlas-b} and CMS \cite{Sirunyan:2017oug}.
}
\end{center}
\end{figure}

The only fitted parameter is $\hat q_0$, which is
 the maximal value of the brodening rate (transport coefficient) at $s=\tau=0$ and $t=t_0$. In fact, measurement of this parameter is our goal.
Comparison with ATLAS \cite{atlas-b} and CMS data \cite{Sirunyan:2017oug}  for $B$-meson production (non-prompt $J/\psi$) in lead-lead collisions at $\sqrt{s}=5.02\TeV$ is presented in Fig.~\ref{Bmeson}.
 %%%%%%%%%%%%%
%%%%%%%%%%%%
We see that data are described pretty well, either for $p_T$, or  $N_{part}$ dependences.
The adjusted parameter $\hat q_0$ ranges within $\hat q_0= 0.2-0.25\GeV^2/\fm$.
This magnitude is considerably smaller compared with the values usually measured for light quarks. See discussion below.

\begin{figure}[hbt]
\begin{center}
\includegraphics[width=13 cm]{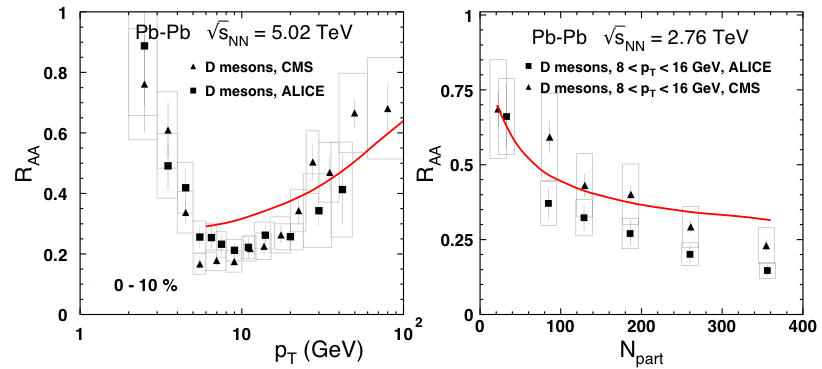}
\caption{\label{Dmeson} The same as in Fig.~\ref{Bmeson}, but for production of $D$-mesons in experiments ALICE \cite{D-alice} and CMS \cite{cms-D-pp}.}
\end{center}
\end{figure}

We successfully described data on $D$-meson production as well, as is demonstrated in 
Fig.~\ref{Dmeson}.
%%%%%%%%%%%%%
%%%%%%%%%%%%
Notice that c-quarks radiate in vacuum more energy than b-quarks, while the effects of absorption of $c\bar q$ and $b\bar q$
dipoles in the medium are similar. Therefore $D$-mesons are suppressed in $AA$ collisions more than $B$-mesons. $R_{AA}(p_T)$ for $D$-mesons steeply rises with $p_T$ due to color transparency. Since $b\bar q$ dipoles expand much faster than $c\bar q$, no color transparency effects are seen in $R_{AA}(p_T)$ for $B$-mesons, as was demonstrated in the right pane of Fig.~\ref{ATLAS}.

Interesting that the found broadening rate parameter for $c$-quarks $\hat q_0= 0.45-0.55\GeV^2/\fm$, significantly exceeds the value  $\hat q_0= 0.2-0.25\GeV^2/\fm$ we found for $b$-quarks, while is quite less than  $\hat q_0\approx 2\GeV^2/\fm$ for light quarks (see below). Such a hierarchy of broadening rates for different quark flavors might look puzzling, if $\hat q$ were a real transport coefficient in terms of statistical medium properties. It coincides with the rate of broadening \cite{bdmps} only within the Born approximation, i.e. single gluon exchange for an inelastic process. In reality, broadening is subject to strong higher-order corrections and usually considerably exceeds the Born approximation estimate. The rate of broadening reads \cite{jkt,mutual},
\beq
\hat q=\frac{2\pi^2}{3}\,\alpha_s(\mu^2)\,xg(x,\mu^2)\,\rho_2,
\label{1057}
\eeq
where $g(x,\mu^2)$ is the gluon density; $\rho_2$ is the medium density per unit of length.
The characteristic scale of the process $\mu$ is related to the mean transverse momentum of the radiated gluons. For light quarks it is given by the non-perturbative effective gluon mass, 
$m_g\sim 0.7\GeV$ \cite{kst2,spots}. For heavy quarks gluon radiation is subject to the dead-cone effect and the scale is much larger $\mu^2\approx m_Q^2$. This is why the rate of broadening for heavy quarks is significantly reduced. This is another manifestation of the dead-cone effect.

\begin{figure}[hbt]
\begin{center}
\includegraphics[width=6 cm]{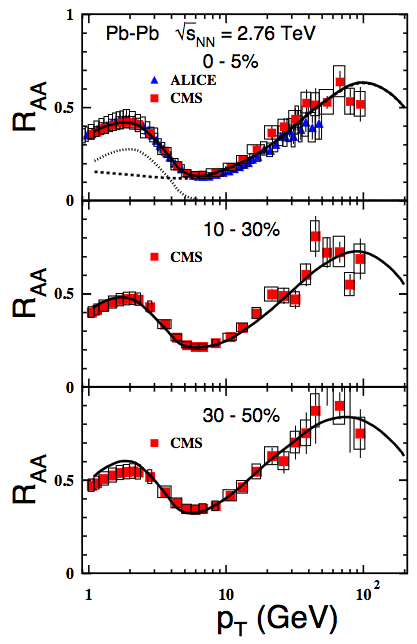}
\caption{\label{Hydro} Suppression factor $R_{AA}(p_T)$ for lead-lead collisions at $\sqrt{s}= 2.76\TeV$ vs centrality. The dashed
and dotted lines are calculated within the pQCD \cite{pert} and hydrodynamic \cite{sinyukov} 
mechanisms, respectively. The solid lines represent both mechanisms summed up. Data for $R_{AA}$ are from the ALICE \cite{alice1} and CMS \cite{cms1,cms2} experiments.}
\end{center}
\end{figure}

The left plot in Fig.~\ref{Dmeson} shows a considerable disagreement with data at small transverse momenta $p_T\lesssim 10\GeV$. While the measured $R_{AA}(p_T)$ is steeply falling with $p_T$,  our calculations predict a nearly constant value. Such kind of disagreement has been observed earlier for light quarks, as is displayed in Fig.~\ref{Hydro}
%%%%%%%%%%%%%
%%%%%%%%%%%%
Apparently a bump at small $p_T$ is presented in $R_{AA}(p_T)$ for $D$-mesons as well, while our calculations in Fig.~\ref{Dmeson} disregard the hydrodynamic component.

\section{Summary}

Our observations and results can be summarized as follows.
\begin{itemize}
\item
Heavy and light quarks originated from hard collisions radiate differently.
The former is subject to the dead-cone effect, suppressing radiation of low-$k_T$ gluons.
Consequently heavy quarks regenerate their color field much faster than light ones
and radiate a significantly smaller fraction of the initial energy. The heavier is a quark, the less it radiates.
\item
The fragmentation function usually depends on two variables $D_{M/q}(z,Q^2)$, fractional light-cone momentum of produced meson, and the scale $Q^2$. However, we consider here the case of "maximal" scale, when the jet energy and the hard scale coincide. This happens e.g. in $e^+e^-$ annihilation, or high-$p_T$ jet production at Feynman $x_F=0$. 

\item
The dead-cone effect suppressing bremsstrahlung of heavy quarks, 
explains the unusual shape of the fragmentation function of heavy quarks $D_{M/Q}(z)$,
observed at LEP and SLAC. It peaks at large fractional momentum $z$, i.e. the produced heavy-light mesons, $B$ or $D$, carry the main fraction of the jet momentum. 
On the contrary, the fragmentation function of light quarks is falling steadily with $z$ towards $z=1$.

\item
Differently from propagation of a small $q-\bar q$ dipole, which survives in the medium due to color
transparency, a $Q-\bar q$ dipole promptly expands to a large transverse size, controlled by  the small mass of the light quark. Such a big dipole has no
chance to remain intact in a hot medium. On the other hand, a breakup of such a dipole
hardly affects the production rate of $Q-\bar q$ mesons.

\item
We successfully described data on $p_T$ and centrality dependence of the production rate of $B$ and $D$ mesons in heavy ion collisions. The only unavoidable parameter of such analyses  is the broadening rate (usually called transport coefficient) of the quark  in the medium. Its maximal value $\hat q_0$ was found $0.2-0.25\GeV^2/fm$, $0.4-0.45\GeV^2/fm$ and $2\GeV^2/fm$ for $b$, $c$ and light quarks respectively. Such hierarchy of the broadening rates is related to the same dead-cone effect.Suppression of bremsstrahlung leads to a considerable reduction of broadening.
\end{itemize}

\acknowledgments{
%\begin{acknowledgments}
 This work was supported in part by grants ANID - Chile FONDECYT 1231062 and 1230391,  by  ANID PIA/APOYO AFB220004, and by ANID - Millennium Science  Initiative Program 
ICN2019\_044.
 The work of J.N. was partially supported by
Grant No. LTT18002 of the Ministry of Education, Youth and Sports of
the Czech Republic, by the project of the European Regional Development
Fund No. CZ.02.1.01/0.0/0.0/16\_019/0000778 and by the Slovak Funding
Agency, Grant No. 2/0020/22.
%\end{acknowledgments}
}

%\reftitle{References}

\end{document}